\documentclass[fleqn,usenatbib]{mnras}

\usepackage{newtxtext,newtxmath}

\usepackage[T1]{fontenc}

\DeclareRobustCommand{\VAN}[3]{#2}
\let\VANthebibliography\thebibliography
\def\thebibliography{\DeclareRobustCommand{\VAN}[3]{##3}\VANthebibliography}

\usepackage{graphicx}	
\usepackage{amsmath}	
\usepackage{amssymb}	

\usepackage{hyperref}
\usepackage{booktabs}
\usepackage{amsmath}
\usepackage{xcolor}

\title[eBOSS ELG GLAM-QPM Mocks]{The Completed SDSS-IV Extended Baryon Oscillation Spectroscopic
Survey: GLAM-QPM mock galaxy catalogs for the Emission Line Galaxy Sample}

\author[Lin et al.]{\parbox{\textwidth}{
Sicheng Lin,$^{1}$\thanks{E-mail: sicheng@nyu.edu}
Jeremy L. Tinker,$^{1}$ 
Anatoly Klypin,$^{2}$
Francisco Prada,$^{3}$ 
Michael R. Blanton,$^{1}$
Johan Comparat,$^{4}$
Kyle S. Dawson,$^{5}$
Arnaud de Mattia,$^{6}$
H\'elion~du~Mas~des~Bourboux,$^{5}$
Will J. Percival,$^{7,8,9}$
Anand Raichoor,$^{10}$
Graziano Rossi,$^{11}$
Alex Smith,$^{6}$
Cheng Zhao$^{10}$
}\\
$^{1}$Center for Cosmology and Particle Physics, Department of Physics, New York University, 726 Broadway, New York, NY 10003, USA\\
$^{2}$Astronomy Department, New Mexico State University, Las Cruces, NM, USA\\
$^{3}$Instituto de Astrof\'{\i}sica de Andaluc\'{\i}a (CSIC), Glorieta de  la Astronom\'{\i}a, E-18080 Granada, Spain\\
$^{4}$Max-Planck-Institut f\"{u}r extraterrestrische Physik (MPE), Giessenbachstrasse 1, D-85748 Garching bei M\"unchen, Germany\\
$^{5}$University of Utah, Department of Physics and Astronomy, 115 S 1400 E, Salt Lake City, UT 84112, USA\\
$^{6}$IRFU, CEA, Universit\'e Paris-Saclay, F-91191 Gif-sur-Yvette, France\\
$^{7}$Waterloo Centre for Astrophysics, University of Waterloo, 200 University Ave W, Waterloo, ON N2L 3G1, Canada\\
$^{8}$Department of Physics and Astronomy, University of Waterloo, 200 University Ave W, Waterloo, ON N2L 3G1, Canada\\
$^{9}$Perimeter Institute for Theoretical Physics, 31 Caroline St. North, Waterloo, ON N2L 2Y5, Canada\\
$^{10}$Institute of Physics, Laboratory of Astrophysics, Ecole Polytechnique Fédérale de Lausanne (EPFL), Observatoire de Sauverny, 1290 Versoix, Switzerland\\
$^{11}$Department of Physics and Astronomy, Sejong University, Seoul 143-747, Korea
}

\date{Accepted XXX. Received YYY; in original form ZZZ}

\pubyear{2020}

\begin{document}
\label{firstpage}
\pagerange{\pageref{firstpage}--\pageref{lastpage}}
\maketitle

\begin{abstract}
We present 2000 mock galaxy catalogs for the analysis of baryon acoustic oscillations in the Emission Line Galaxy (ELG) sample of the Extended Baryon Oscillation Spectroscopic Survey Data Release 16 (eBOSS DR16). Each mock catalog has a number density of $6.7 \times 10^{-4} h^3 \rm Mpc^{-3}$, covering a redshift range from 0.6 to 1.1. The mocks are calibrated to small-scale eBOSS ELG  clustering measurements at scales of around 10 $h^{-1}$Mpc. The mock catalogs are generated using a combination of GaLAxy Mocks (GLAM) simulations and the Quick Particle-Mesh (QPM) method. GLAM simulations are used to generate the density field, which is then assigned dark matter halos using the QPM method. Halos are populated with galaxies using a halo occupation distribution (HOD).  The resulting mocks match the survey geometry and selection function of the data, and have slightly higher number density which allows room for systematic analysis. The large-scale clustering of mocks at the baryon acoustic oscillation (BAO) scale is consistent with data and we present the correlation matrix of the mocks. 
\end{abstract}



\section{Introduction}
\label{sec: introduction}In modern cosmology, the study of the large-scale structure (LSS) provides key information about the expansion history and growth of structure in the Universe \citep{Davis1983,Eisenstein2005}. Measurements of baryon acoustic oscillations \citep[BAO;][]{Eisenstein1998} and redshift space distortions \citep[RSD;][]{kaiser1987} require spectroscopic surveys to cover large volumes and to have accurate redshift measurements. In the past, large redshift surveys have included 2dFGRS \citep{Cole2005}, the Sloan Digital Sky Survey \citep[SDSS-II; ][]{Eisenstein2005}, 6dFGRS \citep{Beutler2012}, WiggleZ  \citep{blake2011} and SDSS-III Baryon Oscillation Spectroscopic Survey \citep[BOSS;][]{dawson2013}. The recently completed extended Baryon Oscillation Spectroscopic Survey \citep[eBOSS; ][]{Dawson2016} is a five year program of the Sloan Sky Digital Survey \citep[SDSS-IV; ][]{blanton2017} and is aiming at measuring the distance-redshift relation with BAO at the percent-level using various galaxy tracers. 

One important question for big redshift surveys is how to determine the uncertainties in the measurements of cosmological parameters. Simulated mock catalogs can be used to estimate the covariance matrix of galaxy clustering and these errors are propagated  to cosmological parameter uncertainties by integrating over the parameter likelihood function \citep{dodelson2013, percival2014}. This method requires accurate mock catalogs to cover a huge volume and large number density of galaxies as the survey geometry and redshift range grow larger. Determining the uncertainties of a large survey such as eBOSS requires thousands of mock catalogs, which can be computationally expensive. In recent years, several quick methods such as quick particle mesh \citep[QPM;][]{White2014}, effective Zel’dovich approximation \citep[EZmocks;][]{chuang2015}  and PerturbAtion
Theory Catalog generator of Halo and galaxY distributions \citep[PATCHY;][]{kitaura2014}, have been developed to generate mocks efficiently.  EZmocks use an ad-hoc model to populate galaxies directly on the dark matter field which is generated using the Zeldovich approximation. A similar approach is used by PATCHY but using Augmented Lagrangian Perturbation Theory (ALPT) instead, while the QPM  method selects dark matter particles that mimic the statistics of dark matter halos, and then  populate galaxies in dark matter halos using a standard halo occupation approach. 

We present a set of mock catalogues for the eBOSS ELG sample, which have been tuned to reproduce the small-scale clustering measurements. In this work, we focus on the ELG sample in the $0.6<z<1.1$ redshift range. We use the GLAM $N$-body simulations \citep{klypin2018} and the QPM method to construct the ersatz halo catalogs. Galaxies are then populated using the halo occupation distribution (HOD) model to match the two-point statistics of the eBOSS ELG measurements.  We produce a set of 2000 accurate  mock catalogs for the estimation of the covariance matrix. The mock catalogs are calibrated to match the ELG clustering  at 10 $h^{-1}$Mpc scales.

This study is part of a series of papers of the final eBOSS DR16 data and cosmological measurements. The BAO and RSD results are presented in  \citet{LRG_corr} and \citet{gil-marin20a} for luminous red galaxies; for emission line galaxies see \citet{raichoor2020},\citet{tamone20a} and \citet{arnaud2020}; and see \citet{hou20a}, \citet{neveux20a} for quasars. The essential data catalogs are presented in \citet{ross20} and \citet{lyke20}, the $N$-body mocks for systematic errors are presented in \citet{rossi20} and \citet{smith20}, another set of approximate mocks is presented in \citet{zhao20}. The ELG mock challenge result is presented in \citet{elg_mock_challenge}, the ELG HOD analysis is presented in \citet{Avila20}. The measurements of BAO in the Ly-$\alpha$ forest is presented in \citet{2020duMasdesBourbouxH}. Lastly, the cosmological interpretation of the full eBOSS sample can be found in \citet{eBOSS_Cosmology}.

This paper is organized as follows: Section 2 describes the eBOSS ELG sample used in the analysis. In Section 3, we present our small-scale clustering measurements, with systematic corrections, that we use to calibrate our mock catalogs. The procedure of creating mock catalogs is described in Section 4. Section 5 compares the large-scale clustering of the ELG sample and GLAM-QPM mocks, and we present the covariance matrix. Finally we discuss our results in Section 6. In this paper, the distances are measured in units of $h^{-1}$Mpc with the Hubble constant $H_0=100h \textnormal{km s}^{-1}\textnormal{Mpc}^{-1}$. We assume a fiducial $\Lambda$CDM cosmology with parameters $(\Omega_m, h, \Omega_bh^2, \sigma_8, n_s) = (0.307, 0.678, 0.022, 0.828, 0.96)$ for the redshift-distance relationship and mock catalog generation.

\section{Data}
The eBOSS survey was conducted using the Sloan Foundation 2.5-m Telescope at Apache Point Observatory \citep{gunn2006} to conduct spectroscopic observations. It used the same 1000-fiber spectrographs as BOSS \citep{Smee2013} to measure four tracers of the underlying dark matter density field. In data release 16, the survey has measured accurate redshifts of 174,816 luminous red galaxies (LRGs) in the  redshift range $0.6<z<1.0$ \citep{ross20}, 173,736 emission line galaxies (ELGs) in the redshift range $0.6<z<1.1$ \citep{raichoor2020}, 343,708 quasars within $0.8<z<2.2$ \citep{ross20, lyke20} and 210,005 Ly$\alpha$ quasars within $z>2.1$ \citep{2020duMasdesBourbouxH}. Overall these datasets constrain the redshift-distance relation to $1\%-3\%$ level at 4 redshifts \citep{Dawson2016,eBOSS_Cosmology}.

The eBOSS/ELG program started in September 2016 and it is the first time ELG tracers have been used in SDSS for large-scale clustering measurements. Preliminary work tested the feasibility of using the BOSS spectrograph to conduct ELG observations \citep{comparat2013} and the reliability of redshift measurements \citep{comparat2015,comparat2016}. The ELG target selection \citep{raichoor2017} uses the DECam Legacy Survey \citep[DECaLS; ][]{Dey2018}. Its deep imaging data provides the opportunity to reach higher redshift and higher efficiency,  defined as percentage of observed ELGs having reliable $z_{spec}$ \citep{raichoor2017}.  The eBOSS/ELG footprint covers an effective area of 369.5 deg$^2$ in the north galactic cap (NGC) and 357.5 deg$^2$ in the south galactic cap (SGC), with an overall number density of 313.0 deg$^{-2}$  \citep{raichoor2020}. 

\begin{figure*}
\begin{center}
\includegraphics[width=0.95\textwidth]{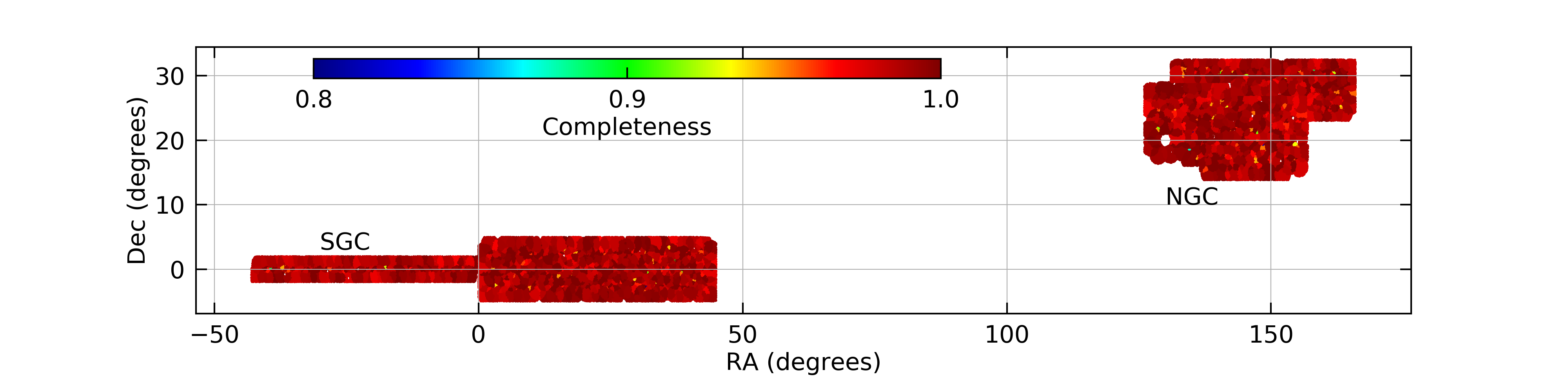}
\caption{The footprint of the ELG SGC and NGC survey, where the color represents the completeness in each sector. The mean completeness in each sector is 0.9909 for NGC and 0.9906 for SGC.}
\label{footprint}
\end{center}
\end{figure*}

We use the eBOSS DR16 ELG sample as our dataset. This dataset includes 83,769 galaxies in SGC and 89,967 galaxies in NGC with reliable redshift measurements over the redshift range $0.6<z<1.1$. Fig. \ref{footprint} shows the footprint of ELG SGC and NGC data colored by completeness in each sector, where a sector is the geometric region defined by the unique set of overlapping plates. The completeness in each sector is defined as 

\begin{equation}
    C = \frac{N_{\rm obs} + N_{\rm cp}}{N_{\rm targ}},
\end{equation}
where $N_{\rm obs}$ is the number of observed targets per sector, $N_{\rm cp}$ is the number of galaxies with no spectra due to fiber collisions with other targets, and $N_{\rm targ}$ is the total number of targets.

\section{Galaxy clustering}\label{sec: clustering}
Galaxies are not randomly distributed in the universe. Density perturbations, which are created in the early Universe, evolve under gravitational attraction, and are the seeds of the large-scale structure we see today. In this paper, study the statistics of the ELG galaxy distribution in the eBOSS survey using the two-point correlation function, $\xi(\textbf{r})=\langle\delta(\textbf{x})\delta(\textbf{x}+\textbf{r})\rangle$, which is a measure of the excess probability of finding a pair of galaxies, separated by a distance $\textbf{r}$, compared to if the galaxies were distributed randomly  \citep{Peebles1980}. Measurements of clustering on large scales allow us to constrain cosmological parameters via measuring the position of BAO peak and the shape of the clustering signal. On small scales, clustering measurements can be used to probe the relationship between galaxy properties and dark matter halos. Thus, we focus on small-scale clustering to calibrate the adopted HOD model to generate the ELG mocks. We will discuss the HOD model further in Sec. \ref{sec:hod}.

We use the Landy \& Szalay estimator \citep{Landy1993} to measure the two-point correlation function, 
\begin{equation}
	\xi(r) = \frac{DD(r)-2DR(r)+RR(r)}{RR(r)},
\end{equation}
where $DD(r), DR(r)$ and $RR(r)$ are suitably normalized numbers of galaxy-galaxy, galaxy-random, and random-random pairs in each distance separation bin. The distance along the line-of-sight of galaxies is inferred from their redshifts assuming a fiducial cosmological model. The peculiar velocities of galaxies will introduce redshift-space distortions in $\xi(r)$. In order to circumvent the effect of redshift-space distortions, the correlation function is often measured in two dimensions: perpendicular ($r_p$) and along ($\pi$) the line-of-sight. Let $\textbf{v}_1$ and $\textbf{v}_2$ be the position vector of a pair of galaxies in redshift-space, $\textbf{s}=\textbf{v}_1-\textbf{v}_2$ be the redshift space separation, and $\textbf{l}=(\textbf{v}_1-\textbf{v}_2)/2$ be the mean position of galaxy pair. The distances $\pi$ and $r_p$ can then be defined as

\begin{equation}
	\pi=\frac{\textbf{s}\cdot \textbf{l}}{|\textbf{l}|},~~~~r_p=\sqrt[]{\textbf{s}\cdot\textbf{s}-\pi^2}.
\end{equation}
One can then measure $\xi(r_p, \pi)$ from the data and random catalogs. The projected correlation function \citep{Davis1983} can be recovered by integrating over the line-of-sight direction to remove the effect of RSD, 
\begin{equation}
	w_p(r_p) = 2\int_0^{\pi_{\rm max}} \xi(r_p, \pi) d\pi.
\end{equation}
In this paper, we choose $\pi_{\rm max}=80 h^{-1}\rm Mpc$ as the upper limit in the integral, as the clustering measurements are noisy for large $\pi$. The correlation function $\xi(r_p, \pi)$ can be alternatively expressed as a function of $s$ and $\mu$, where $\mu = \cos\theta$ is the cosine of the angle between the pair separation vector and line of sight. It is often useful to compress the redshift space information in the two-dimensional correlation function into the multipole moments,
\begin{equation}
	\xi_l(s) = \frac{2l+1}{2} \int \xi(s, \mu) P_l(\mu) d\mu,
\end{equation}
where $P_l(\mu)$ are Legendre polynomials.

We consider the effects of imaging systematics, redshift failures and fiber collisions. These are corrected for using a weighting scheme similar to \citep{anderson2014}, where each ELG is weighed by
\begin{equation}
	w_{\textnormal{ELG}} = w_{\textnormal{FKP}}w_{\textrm{sys}}w_{\textrm{cp}}w_{\textrm{noz}},
\end{equation}
where $w_{\textnormal{FKP}}$ is the FKP weight \citep{fkp1994}, $w_{\textrm{sys}}$ is the imaging systematics weight, $w_{\textrm{cp}}$ is the close-pair weight and $w_{\textrm{noz}}$ is the redshift failure weight. We describe them in detail below. Unlike \citet{anderson2014}, here we treat $w_{\textrm{cp}}$ and $w_{\textrm{noz}}$ independently.

The systematic weights, $w_{\textrm{sys}}$, are calculated by using a linear fit of the ELG target density in various photometric variables: galactic extinction, stellar density, HI density, $grz$-band image seeing, and $grz$-band image depth \citep{raichoor2020}. The close-pair weights, $w_{\textrm{cp}}$, are calculated by upweighting galaxies in collided pairs
by coefficient $N_{\rm target}/N_{\rm fiber}$, where $N_{\rm target}$ is the total number of targets in the collision group and $N_{\rm fiber}$ is the number of targets in the collision group that has been assigned fibers. The redshift failure weights, $w_{\textrm{noz}}$, are the inverse of two fitting functions  to the  plate-averaged signal-to-noise ratio (pSN) and the fiber position in the focal plane.  Details of the systematic model is described in the eBOSS ELG catalog paper \citep{raichoor2020}.

The FKP weights are defined as
\begin{equation}
	w_{\textrm{FKP}}(z) = \frac{1}{1+n(z)P_{\textrm{FKP}}},
\end{equation}
where $n(z)$ is the weighted number density at redshift $z$ and we choose the same value of $P_{\textrm{FKP}}=4000h^{-3}\textnormal{Mpc}^3$ as in \citet{raichoor2020}. We use FKP weights to account for different number densities of observed ELGs in different redshift intervals.

In addition to using the close pair weights to up-weight galaxies in collided pairs, we also apply an angular weighting to correct the small-scale clustering measurements, using the ratio of the angular correlation functions \citep{hawkins2003}. This ratio is given by
\begin{equation}
	F(\theta) = \frac{1+w_z(\theta)}{1+w_t(\theta)},
\end{equation}
where $w_z(\theta)$ is the angular correlation function of galaxies with fibers assigned and $w_t(\theta)$ is the angular correlation function of the parent target samples. We find that the average ratio $F(\theta)$ for angular separations $\theta<62''$ is around $0.6$ and is not sensitive to the value of $\theta$, thus we upweight each galaxy-galaxy pair below $62''$ by $1/F(\theta)=1.667$.

\begin{figure}
\begin{center}
\includegraphics[width=\columnwidth]{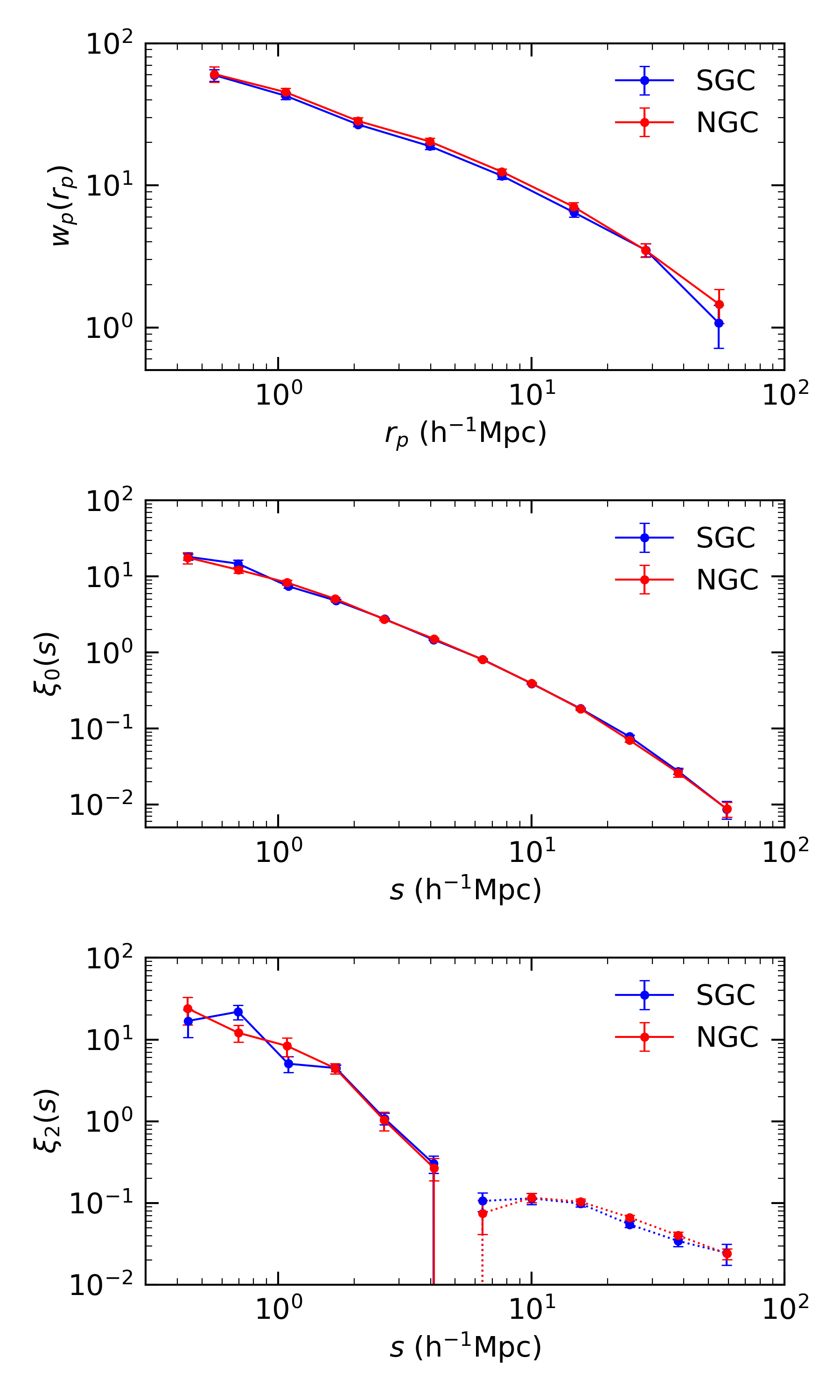}
\caption{Projected correlation function, $w_p$ (upper panel), redshift-space monopole, $\xi_0$ (middle panel) and quadrupole, $\xi_2$ (lower panel) for ELGs in the SGC (blue) and NGC (red). Correlation functions are shown for separations between $0.3h^{-1}\rm Mpc$ and $70h^{-1}\rm Mpc$, for ELGs in the redshift range $0.6<z<1.1$. Dotted curves in the lower panel indicate negative values. The errorbars are estimated using jackknife resampling, with 25 jackknife samples.}
\label{monopole}
\end{center}
\end{figure}

Fig. \ref{monopole} presents our results of the small-scale clustering from $0.34h^{-1}\rm Mpc$ to $70h^{-1}\rm Mpc$ in 12 logarithmic bins, after applying the weighting described above. The SGC and NGC are divided into 25 roughly equal spherical areas and the errors are estimated using the jackknife resampling technique.  There is no significant difference seen between the clustering of the SGC and NGC.

\section{Mock generation}

In order to construct accurate mocks to interpret the ELG clustering, we combine GaLAxy Mocks (GLAM) simulations with the Quick Particle-Mesh (QPM) scheme. The whole process can be summarized in the following steps:

\begin{itemize}
\item In the first step we run 2000 large GLAM N-body simulations with box size of $3000 h^{-1}\textnormal{Mpc}$. This box size is large enough to cover the whole footprint of ELGs up to redshift $z=1.1$.
\item In the second step we apply the QPM code to assign dark matter halos within the density field of the GLAM simulations.
\item In the third step we populate the halos with galaxies using a HOD model that is calibrated to reproduce the small-scale clustering measurements of the data.
\item In the fourth step, we cut the mock catalogues according to the ELG survey geometry. We compare the large-scale clustering of the mocks with the data.
\end{itemize}

In the next sub-sections, we will describe the details of generating the mock galaxy catalogs.

\subsection{GLAM simulations}
GLAM \citep{klypin2018} is a new parallel version of the Particle-Mesh (PM) code that can quickly produce a large number of N-body simulations. We use the GLAM code to generate the matter density field for our mock catalogs since the computational speed is much faster than for QPM simulations. We used MareNostrum-4 computer at Barcelona Supercomputer Center to generate 2000 realizations in the adopted cosmology (see Sec. \ref{sec: introduction}). The volume of each simulation is $3h^{-1}\rm Gpc$, which is large enough to cover the ELG redshift range $0.6<z<1.1$. We used $1500^3$ particles with a mass per particle of $6.8\times10^{11}h^{-1}\rm M_\odot$. The simulations were started at $z=100$ and a constant time-step is used at low redshifts but periodically increases at high redshifts. The total number of time steps is 94, which is large enough to satisfy both accuracy and stability of the integration of the particle trajectories inside dense regions \citep{klypin2018}. Under this set of simulation parameters, the total number of CPU hours is 52,000. In order to make the process as efficient as possible, we incorporate QPM as a module in the GLAM code, so halo catalog creation is done on the fly. This procedure prevents extra time consumption due to I/O of large files, and saves disk storage.

\subsection{From N-body simulation to halo catalogs} 
There are many benefits to first create halo catalogs  and then generate galaxy mocks using galaxy-halo models. One of which is that we can model multiple target samples with different biases within the halo occupation framework. The other advantage is that we can study and test different galaxy-halo models for the same sample. 

\begin{figure}
\begin{center}
\includegraphics[width=\columnwidth]{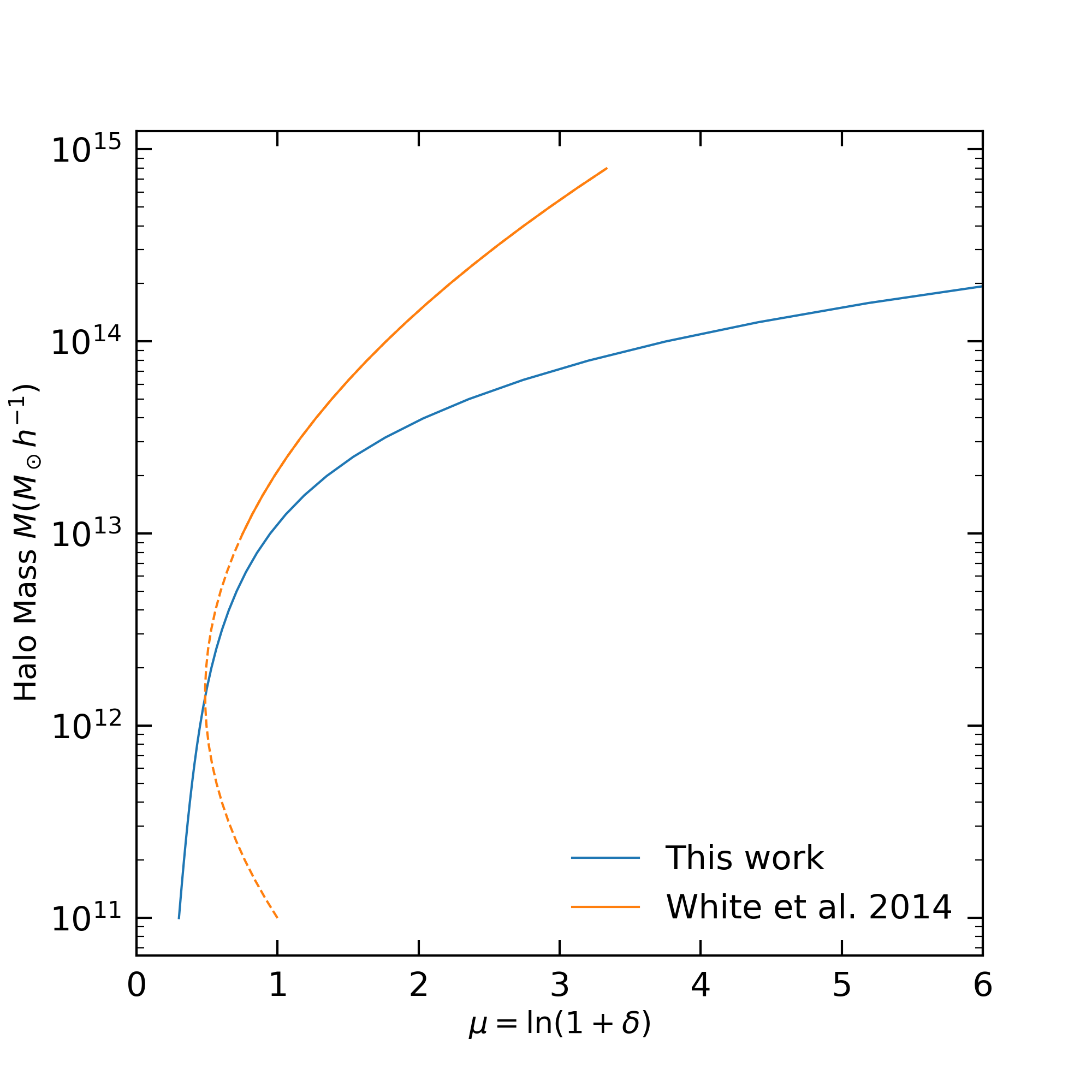}
\caption{Fitting function describing the relationship between the dark matter overdensity and halo mass, where $\mu=\ln(1+\delta)$. The fitting function used in this work (blue) is approximately logarithmic approximation at low halo masses, transitioning to a power law at the high halo mass end. For comparison, the mapping used in \citet{White2014} is shown in orange. The scatter of the mapping is 0.1 dex, as described in Eq. \ref{eq2}. Only halos with $M_h>10^{13}h^{-1}\textrm{M}_\odot$ are used in \citet{White2014}. The functions differ at high masses since they are for halos at different redshifts.}
\label{den_mass}
\end{center}
\end{figure}

Here we use a modified version of the QPM method described in \citet{White2014} to generate halo catalogs from the GLAM simulations. First, we use Fourier methods on a mesh grid to estimate the density field of the GLAM simulations, which is then mapped to a halo mass. We calibrate the mapping scheme so as to match the bias of halos from high resolution simulations \citep{Tinker2008}.  We group particles in the GLAM simulations by their density $\mu = \ln(1+\delta)$ in 8 equally spaced bins, then calculate the bias of each group. The bias is then mapped to a halo mass using the halo bias function, $b(M_h)$, of \citet{tinker2010}, We then fit a smooth function to $\mu(M_h)$ of the particles. The result fitting function is
\begin{equation}
\label{eq1}
    \mu(M_h) = 0.5 + 0.1\log_{10}(M_h/M_0) + \frac{(M_h/M_0)^{0.7}}{1+(M_h/M_0)^{-0.35}},
\end{equation}
where $M_h$ is the halo mass and $M_0 = 10^{13.5}h^{-1}\rm M_\odot$ is the transition scale from a logarithmic function to power law. The shape of the function is shown in Fig. \ref{den_mass} compared to the function adopted in \citet{White2014}. Note that \citet{White2014} only use halos with $M_h>10^{13}h^{-1}\textrm{M}_\odot$. Compared with \citet{White2014}, our fitting function allows us to resolve low-mass halos down to $10^{11}h^{-1}\textrm{M}_\odot$. This is important for the ELG samples because ELGs are relatively young galaxies and are more likely to reside in smaller halos compared to luminous red galaxies \citep{violeta2018}. Our fitting function differs from \citet{White2014} at high masses because they are at different redshifts. We aim to produce mocks at redshift $z=0.84$ for eBOSS ELG sample, while \citet{White2014} uses redshift $z=0.55$ for the BOSS LRG sample.

Secondly, we select a subset of particles based on the density to stand in for halos, and these halos are assigned the same positions and velocities as the particles.  Particles are sampled using a Gaussian sampling function. A particle with density $\mu_0$ is assigned a halo mass $M_h$ with a probability of 
\begin{equation}
\label{eq2}
P(M_h |\mu_0, \sigma) = \frac{1}{\sqrt{2\pi\sigma^2}}\exp\left(\frac{(\mu(M_h)-\mu_0)^2}{2\sigma^2}\right),
\end{equation} where the  function $\mu(M_h)$ is defined in Eq. \ref{eq1}, and we fix $\sigma=0.1$. The width $\sigma$ doesn't have a significant effect for the large-scale bias, as shown in \citet{White2014}. Since the total number of particles in each simulation is finite, there are not enough particles at low halo masses. This imposes a mass resolution in our halo catalogs at a mass $M_h>2\times10^{11}h^{-1}\textrm{M}_\odot$.

Finally, in order to have the correct halo mass function, we divided the halo masses into $N_h$ bins, where the number $N_h$ is large so that the change in $b(M_h)$ between each bin is small. For each mass bin we calculate the number of particles we need to sample to match the mass function $n(M_h)$ of \citet{Tinker2008}, then we loop over all particles, assigning particles as halos with the corresponding probability.

\begin{figure}
\begin{center}
\includegraphics[width=\columnwidth]{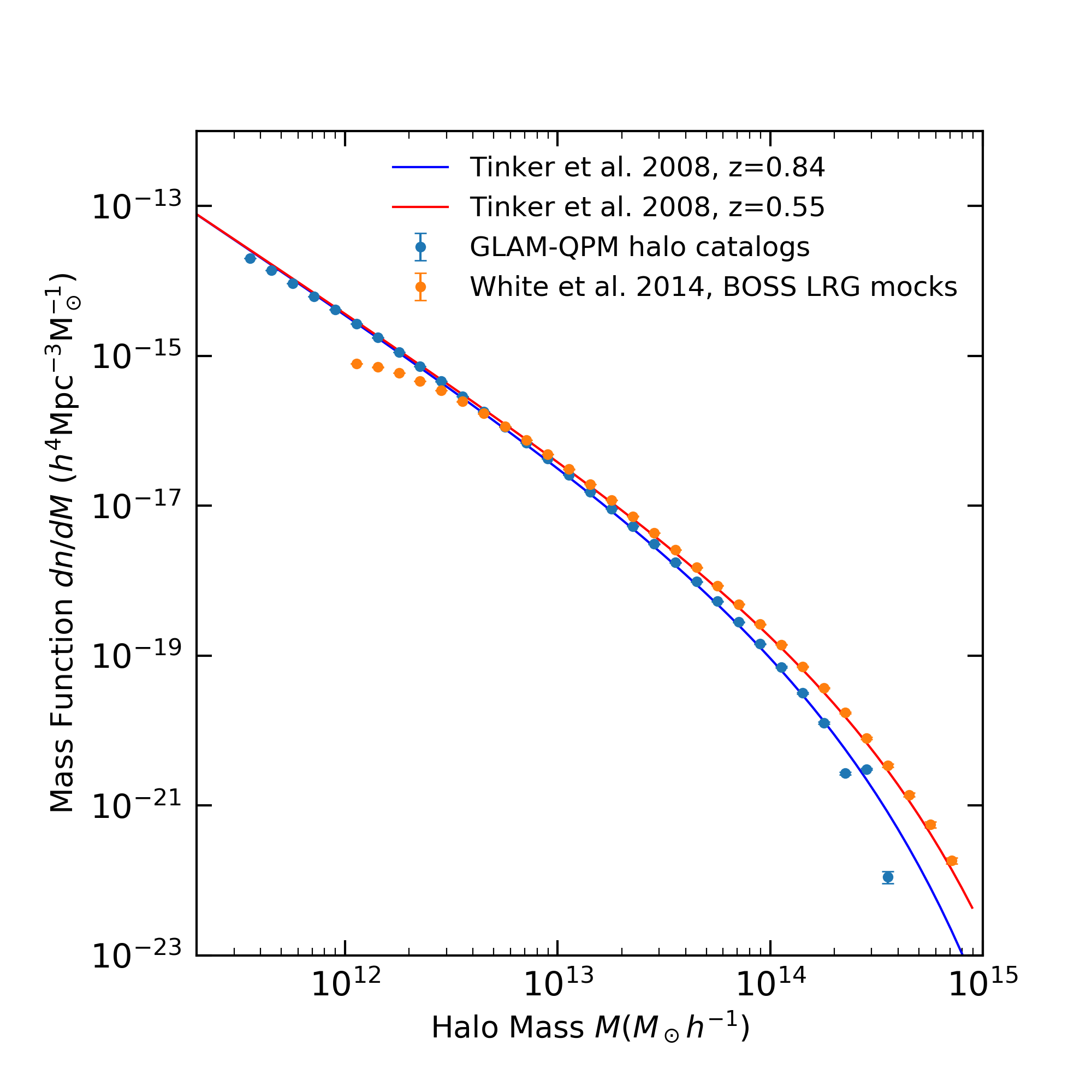}
\caption{The halo mass function from the GLAM-QPM mocks generated in this work (blue points) compared to BOSS LRG mocks (orange points), which are created using the QPM method \citep{White2014}. For both, we show the average mass function of 10 mocks. The errorbars are the standard deviation among the 10 halo catalogs. As a comparison, we also plot the halo mass function from \citet{Tinker2008} at redshift $z=0.55$ (red) and $z=0.84$ (blue).}
\label{hmf}
\end{center}
\end{figure}

In Fig. \ref{hmf}, we show the average halo mass function from 10 halo catalogs. The 10 catalogs are generated independently from different GLAM simulations. Compared to the QPM mocks used for BOSS LRG sample \citep{Alam2017}, our halo catalogs agree very well to the halo mass function in \citet{Tinker2008} at low halo mass. This verifies that our method yields the correct halo mass function. The small discrepancy at the high halo mass end is due to the power law formalism we choose for $\mu(M_h)$, which means it is less likely to have halos with mass larger than $10^{15}h^{-1}\rm M_\odot$. Since the fraction of ELGs in halos with  $M_h\gtrsim10^{15}h^{-1}\rm M_\odot$ is negligible, this discrepancy will not have a significant impact on our mocks.

\begin{figure}
\begin{center}
\includegraphics[width=\columnwidth]{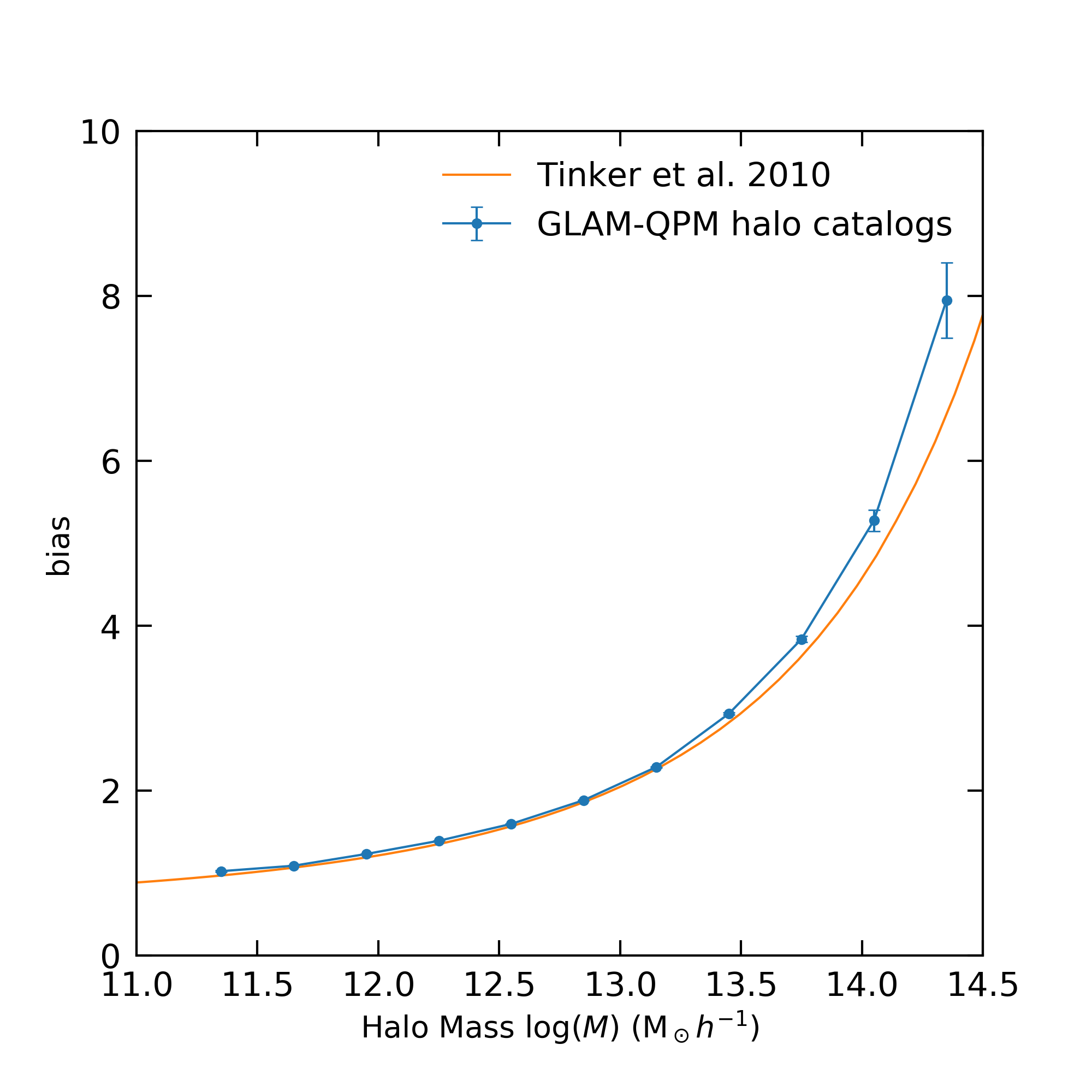}
\caption{The average halo bias from from 10 GLAM-QPM halo catalogs (blue), where the error bars indicate the standard error on the mean. The orange curve indicates the fitting function of \citet{tinker2010} at redshift $z=0.84$. }
\label{halo_bias}
\end{center}
\end{figure}

In Fig. \ref{halo_bias}, we compare the halo bias calculated from the halo catalogs with the halo bias of \citet{tinker2010}. The result are in very good agreement for $M_h<10^{13.5} h^{-1} \rm M_\odot$. The small discrepancy at higher halo mass is due to lack of information of the bias in high density regions: there is insufficient number of particles that have high density to compute the bias. Considering that the mean halo mass of ELGs is around $10^{12} h^{-1} \rm M_\odot $, this discrepancy should be negligible. 

\subsection{Halo occupation distribution for ELGs}\label{sec:hod}

We use the halo occupation distribution (HOD) to model galaxy bias \citep{Peacock2000,Seljak2000,Benson2000,scoccimarro2001,Zheng2009,White2011,Berlind2002}. The HOD formalism describes the relation between a typical class of galaxies and dark matter halos by the probability $P(N|M)$ that a halo with mass $M$ contains $N$ such galaxies. The population of galaxies can be split into central galaxies, which reside at the center of the halo, and satellite galaxies. Here we assume that the satellite galaxies in each halo follow the same radial distribution as the dark matter, corresponding to a NFW profile \citep{NFW1997} where we use the concentration-mass relation of \citet{maccio2007}. The HOD model is a complete description of galaxy bias, i.e., given an HOD model and the halo population from a cosmological model, one can calculate any galaxy clustering statistic on any scale.

HOD modeling has been applied to interpret galaxy clustering in several surveys \citep{Bullock2002,Moustakas2002,vandenBosch2003,zheng2004,zehavi2005,Lee2006,zheng2007,Wake2008,Blake2008,Zheng2009,Parejko2013,Park2016,Zhai2017,Carter2018}. Most of the studies use five parameter HOD models: the occupation function for central galaxies $\langle N_{\rm cen}\rangle$ is a softened step function with three parameters and the occupation function for satellite galaxies $\langle N_{\rm sat}\rangle$ is a power law with two free parameters. However, the step function for central galaxies may not model ELGs well, since one would expect that most of the galaxies in higher mass halos are quenched or have attenuated star formation. Therefore it is less likely that central galaxies with strong emission lines will be found in massive halos. Currently, the halo occupation distribution of ELGs are not well understood. \citet{favole2017} studied the HOD of [OII] emitters in the local universe using the (Sub)Halo Abundance Matching (SHAM) method. \citet{violeta2018} studied the properties of the host halos of [OII] emitters, and they found that the central galaxy occupation $\langle N\rangle_{[\textnormal{OII}]\rm cen}$ can be formalized as the sum of a Gaussian and a step function with amplitude below unity. \citet{zehavi2011} studied color dependence of galaxy clustering by fitting an HOD model to red/blue galaxy populations of the SDSS DR7 main galaxy sample, and their central galaxy occupation function is modeled as the difference between two softened step functions.  Based on the results of previous works, we use a Gaussian function with three free parameters for central galaxies and a power law for satellite galaxies,

\begin{equation}
	\langle N_{\rm cen}\rangle = f_{\rm max}\times\exp\left[-\frac{(\log M-\log M_{\rm min})^2}{2\sigma_{\log M}^2}\right],
\end{equation}
\begin{equation}
	\langle N_{\rm sat}\rangle = 
    \left\{
    	\begin{array}{lr}
    		0,& M<M_{\rm cut} \\
            \left(\frac{M-M_{\rm cut}}{M_1}\right)^\alpha,& M \geq M_{\rm cut}\\
    	\end{array}
    \right.
\end{equation} where $M_{\rm min}$ is the characteristic halo mass that has the maximum probability of hosting a central galaxy, $\sigma_{\log M}$ is the standard deviation width in log mass and $f_{\rm max}$ is the maximum probability that a halo host a central galaxy. For satellite galaxies, $\alpha$ is the power law slope,  $M_1$ is the amplitude and  $M_{\rm cut}$ is the cut-off mass.

This formalism is simplified from previous studies, but it is sufficient to model the ELG clustering. The six free parameters in our HOD model are $\{f_{\rm max}, M_{\rm min}, \sigma_{\log M}, M_{\rm cut}, M_1, \alpha\}$. The units of $M_{\rm min}, M_1$ and $M_{\rm cut}$ are $h^{-1} \rm M_\odot$, while $\alpha$, $f_{\rm max}$ and $\sigma_{\log M}$ are dimensionless quantities. We perform a coarse HOD parameter space search for parameter optimization. A value for each parameter is selected for $\{M_{\textrm{min}},M_{\textrm{cut}}, M_1, \alpha\}$ while fixing $f_{\rm max}=0.15$ and $\sigma_{\log M}=0.25$. The parameter space is shown in Table \ref{parameter_table} .
\begin{table}[h]
\begin{center}
\caption{HOD parameter space for parameter optimization.}
\begin{tabular}{@{}ll@{}}
\toprule
           & parameter space\\ \midrule
$M_{\rm min}$ $[h^{-1}\rm M_\odot]$      & $1.0\times10^{12}, 1.5\times10^{12}, 2.0\times10^{12}$ \\
$M_1$   $[h^{-1}\rm M_\odot]$      & $2.0\times10^{13},5.0\times10^{13},8.0\times10^{13}$   \\
$M_{\rm cut}$  $[h^{-1}\rm M_\odot]$   & $7\times10^{11}, 4\times10^{12}$   \\
$\alpha$      & $0.8, $1.0$,1.2$     \\ \bottomrule
\end{tabular}
\label{parameter_table}
\end{center}
\end{table}
In subsequent sections we will demonstrate that the ELG clustering is relatively insensitive to the choice of parameters. Thus, we did not explore the whole HOD parameter space, which would be out of the scope of this paper. For the purpose of mock generation, our method is sufficient to model the large-scale bias of ELGs, with reasonable choices for the satellite fraction of the target sample. 

There are 54 sets of parameter in total. For each set, we generate a galaxy mock and measure the redshift-space monopole $\xi_0(s)$, quadrupole $\xi_2(s)$, as well as projected correlation function $w_p(r_p)$. We calibrate the HOD by finding the set of parameters which most closely match the clustering of the data at scales between 10 $h^{-1}$Mpc and 30 $h^{-1}$Mpc, as one would interpret the linear bias at this scale. At smaller scales, i.e. $s\sim 1 h^{-1}$Mpc, the QPM scheme does not model the bias well. We will illustrate this point presently. At larger scales, the uncertainty in the clustering measurements from the data is large due to sample variance, and the clustering measurements are more prone to residual photometric systematics. The best fit HOD parameters that we use to generate ELG mocks are listed in Table \ref{hod_table}. We tested the effect of varying $f_{\rm max}$ and $\sigma_{\log M}$, and found that this does not significantly change the clustering measurements.

\begin{table}[h]
\begin{center}
\caption{HOD parameters for our fiducial galaxy mocks.}
\begin{tabular}{@{}ll@{}}
\toprule
           & best-fit\\ \midrule
$M_{\rm min}$       & $1.5\times10^{12}~ h^{-1}\rm M_\odot$ \\
$M_1$         & $8\times10^{13}~ h^{-1}\rm M_\odot$   \\
$M_{\rm cut}$     & $7\times10^{11}~ h^{-1}\rm M_\odot$   \\
$\alpha$      & $1.0$     \\
$f_{\rm max}$        & $0.15$    \\
$\sigma_{\log M}$ & $0.25$    \\ \bottomrule
\end{tabular}
\label{hod_table}
\end{center}
\end{table}

\begin{figure*}
\begin{center}
\includegraphics[width=0.95\textwidth]{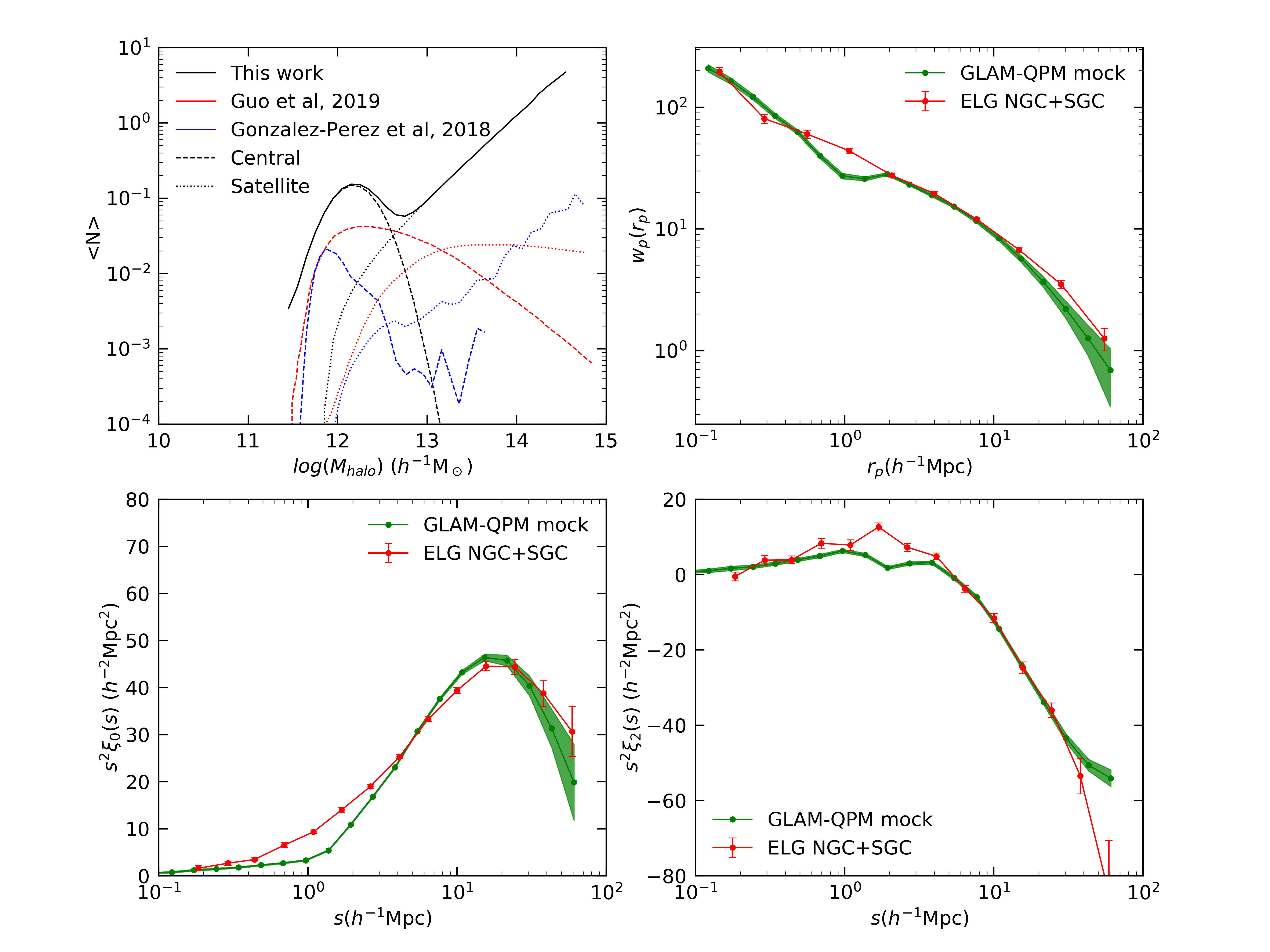}
\caption{HOD model and galaxy clustering comparison between the GLAM-QPM mocks and ELG data. The upper left panel shows the best fit HOD model for the eBOSS ELG sample (black), where the dashed curve indicates the central HOD, dotted curve the satellite HOD and the solid curve is the total HOD.  The satellite fraction of our HOD model is $17.4\%$. As a comparison, we also show the HOD prediction for eBOSS ELG in the redshift range $0.8<z<0.9$ from \citet{violeta2018} (blue) and \citet{guo2019} (red). The remaining three panels show the measurements of $w_p(r_p)$,  $\xi_0(s)$ and $\xi_2(s)$, respectively. The red curves are measurements from ELG data, averaged between the SGC and NGC, weighted by area. The errorbars are estimated using the jackknife resampling technique, with 25 jackknife samples. The green curves are the mean of the clustering of 100 GLAM-QPM mocks using the best fit HOD parameters in Table \ref{hod_table}. The errorbars are the standard deviation, indicating the $1\sigma$ scatter. }
\label{mock_data_compare}
\end{center}
\end{figure*}

The occupation function of central galaxies and satellite galaxies is shown in the upper left panel of Fig. \ref{mock_data_compare}. The satellite fraction is $17.4\%$, in accordance with $\sim 20\%$ satellite fraction of star forming galaxies in \citet{tinker2013}. This is also in agreement with \citet{favole2016}, where it was found that $22.5\pm2.5\%$ of ELGs at redshift 0.8 are satellite galaxies. The galaxy number density produced by our HOD is $6.7\times 10^{-4} h^3 \rm Mpc^{-3}$, which is slightly higher than the peak ELG number density $6.4\times 10^{-4} h^3 \rm Mpc^{-3}$. This is intended to be so, as the higher number density allows studies of systematic corrections. This HOD implies that ELGs reside in halos with mass larger than $10^{11} h^{-1} M_\odot$, and most of the central galaxies in halos with mass larger than $ 10^{13}h^{-1} M_\odot$ are quenched, which is in agreement with the HOD analysis at redshift $z\sim 0.85$ of \citet{tinker2013}. We also compare our result with the HOD study of eBOSS ELGs in \citet{violeta2018} and \citet{guo2019} at redshift $0.8<z<0.9$. The amplitude of our HOD function is higher than the results of \citet{violeta2018} and \citet{guo2019} because we aim to match the peak number density of eBOSS ELG targets for the purpose of producing mock catalogs. In addition their results are for ELGs in the redshift bin $0.8<z<0.9$, while we cover a wider redshift range. Fig. \ref{mock_data_compare} also compares the clustering measurements between mocks and data. The clustering is measured from 100 mocks generated with the HOD parameters in Table \ref{hod_table}. The errorbars represent the $1\sigma$ scatter between mocks. The mocks agree with data at scales from a few $h^{-1}$Mpc to around 20 $h^{-1}$Mpc. At scales around $s\sim 1 h^{-1}\textnormal{Mpc}$, the mocks are less clustered than the data. The reason is that the method of sampling the density field to find halos does not yield the correct number of halos with small separations, thus leading to a deficit in the correlation function at the transition scale between one-halo and two-halo galaxy pairs. The lower halo mass region of the ELG sample accentuates this effect relative to earlier mocks with luminous red galaxies. Due to this fact, we don't perform a $\chi^2$ test on different HOD parameters because we don't want the clustering measurements at small scale $(s\sim 1 h^{-1}\textnormal{Mpc})$ have undue influence on the selection of HOD parameters.

\begin{figure*}
\begin{center}
\includegraphics[width=0.8\textwidth]{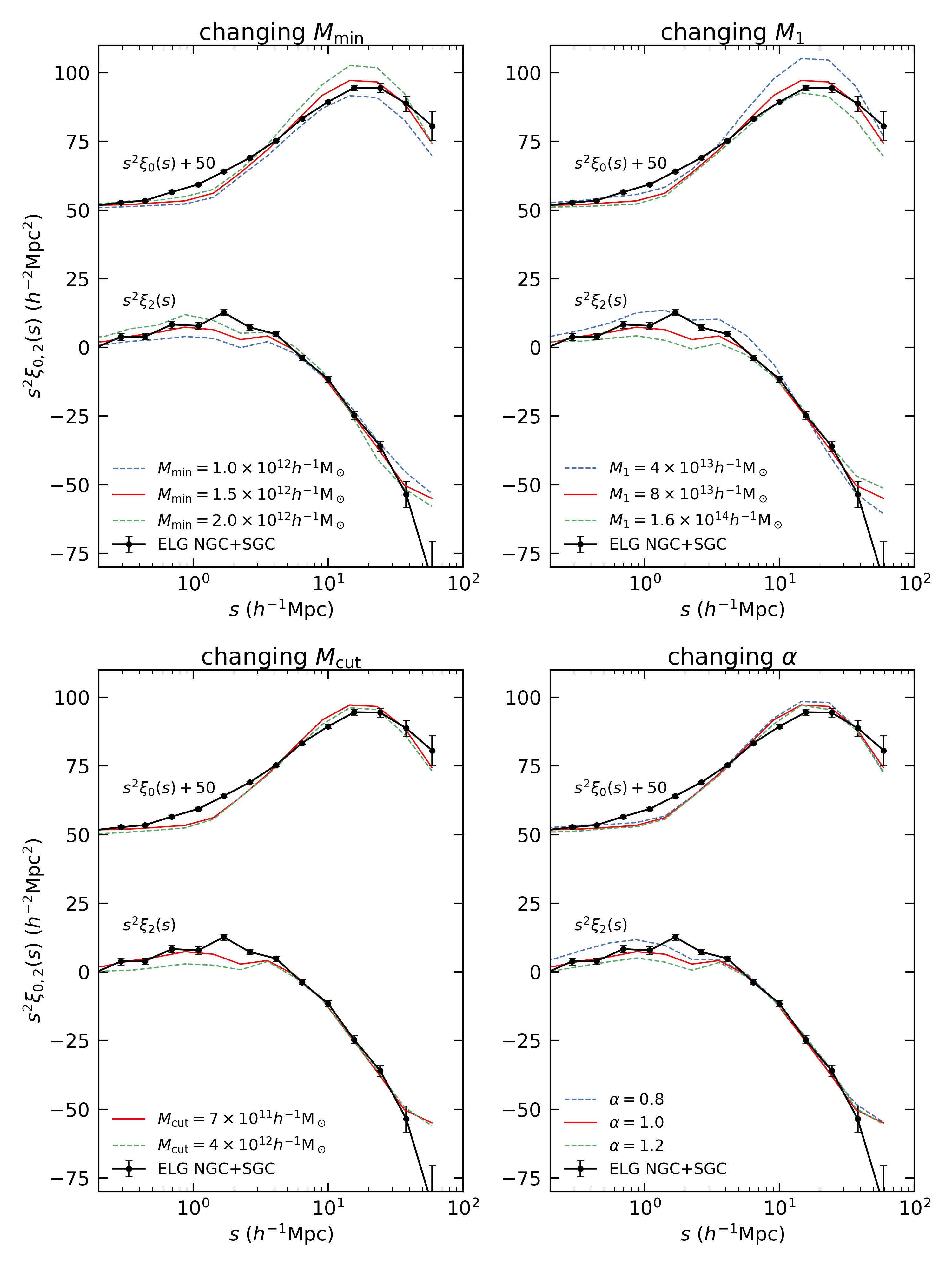}
\caption{The impact of varying the fiducial HOD parameters on the redshift-space monopoles and quadrupoles. Redshift-space monopoles are vertically shifted by 50 to conveniently visualize. Our results for the fiducial set of HOD parameters (see Table \ref{hod_table}) are shown with red curves, and each panel shows the result of perturbing one parameter at a time. The upper left panel shows the impact of varying $M_{\rm min}$, and the upper right panel shows the impact of varying $M_1$. The lower panels show the impact of varying $M_{\rm cut}$ and $\alpha$ on the ELG clustering. The measured clustering of the eBOSS ELGs is indicated by the black points. }
\label{HOD_test}
\end{center}
\end{figure*}

Instead, we test the impact of HOD parameters to the small-scale clustering of ELGs by changing one parameter around the best-fit parameter set. As shown in Fig. \ref{HOD_test}, we fix $f_{\rm max}=0.15, \sigma_{\log M}=0.25$ and perturb other parameters around our fiducial parameter set in Table \ref{hod_table}. The red curve shows the fiducial HOD parameter set, the blue and green curves show the clustering after the perturbation. $M_{\rm min}$ and $M_{1}$ affect the clustering of ELGs the most. This is because $M_{\rm min}$ is the mean halo mass of the Gaussian function for central galaxies, so this parameter determines the halo mass scale that central galaxies reside in, and determines the ELG linear bias. $M_1$ affects the mass of halos that satellite galaxies reside in, so a smaller $M_1$ would result in satellite ELGs being placed in higher mass haloes, increasing the linear bias. There is no significant change in the clustering even if we increase $M_{\rm min}$ and $M_{1}$ by a factor of 2. The impact of $M_{\rm cut}$ and $\alpha$ is even less. We conclude that by choosing suitable HOD parameters, we can produce mock catalogs with small-scale clustering that is in reasonable agreement with the data, and the large-scale bias is relatively insensitive to the parameters chosen.

\section{Large Scale Clustering and Covariance Matrix}

\subsection{Large Scale Clustering}
We generate 2000 galaxy mocks with the HOD parameters presented in Table \ref{hod_table}, and cut the mock to the ELG chunk geometry as well as redshift distribution $n(z)$. We measure the redshift-space monopole and quadrupole up to $200 h^{-1}\textnormal{Mpc}$ using the same method and systematic weights as described in Sec. \ref{sec: clustering}. We choose 40 equally spaced bins for $s$ from $0h^{-1}$Mpc to $200h^{-1}$Mpc. 

\begin{figure}
\begin{center}
\includegraphics[width=\columnwidth]{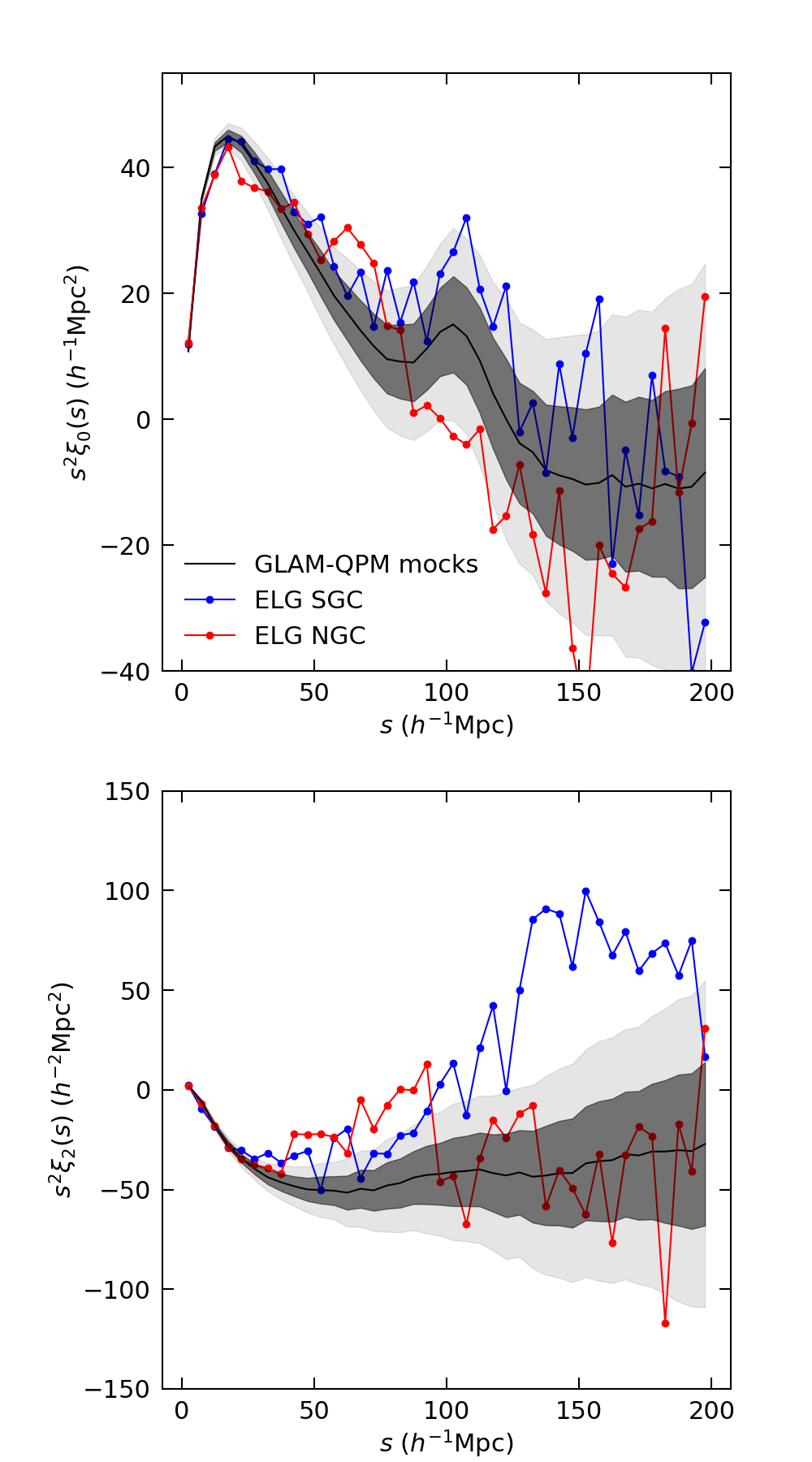}
\caption{Large-scale clustering of ELG data and mocks. The upper panel shows redshift space monopoles and lower panel shows redshift space quadrupoles. The ELG SGC and NGC clustering are colored in blue and red, respectively. The clustering of GLAM-QPM mock is shown in black, with the gray region showing the $1\sigma$ and $2\sigma$ range. }
\label{xi_LSS}
\end{center}
\end{figure}

The large-scale clustering of the mocks is shown in Fig. \ref{xi_LSS}, in comparison with the data. The shaded area indicates the $1\sigma$ and $2\sigma$ scatter in the 2000 mocks. The BAO feature can be clearly seen in the mocks, but there is some visible discrepancy between the mocks and data at the BAO scale, and also between the SGC and NGC. We first present the covariance matrix in Sec. \ref{subsec: covmatrix}, then test the statistical significance of the differences between SGC and NGC in Sec. \ref{subsec: NGC_SGC_diff}.

\subsection{Covariance Matrix}
\label{subsec: covmatrix}
Given a set of mock catalogs, the covariance matrix is defined as:
\begin{equation}
	\textnormal{Cov}_{ij} = \frac{\sum_{k=1}^{n_s} (x_i^k-\mu_i)(x_j^k-\mu_j)}{n_s-1},
\end{equation}
where $x_i$ is the clustering measurement at the $i^{\textrm{th}}$ bin; index k indicates $k^{\textrm{th}}$ realization of mocks; $n_s$ is the total number of mocks; i, j are bins of separation; $\mu_i$ is the mean of $x_i^k$. 

\begin{figure}
\begin{center}
\includegraphics[width=\columnwidth]{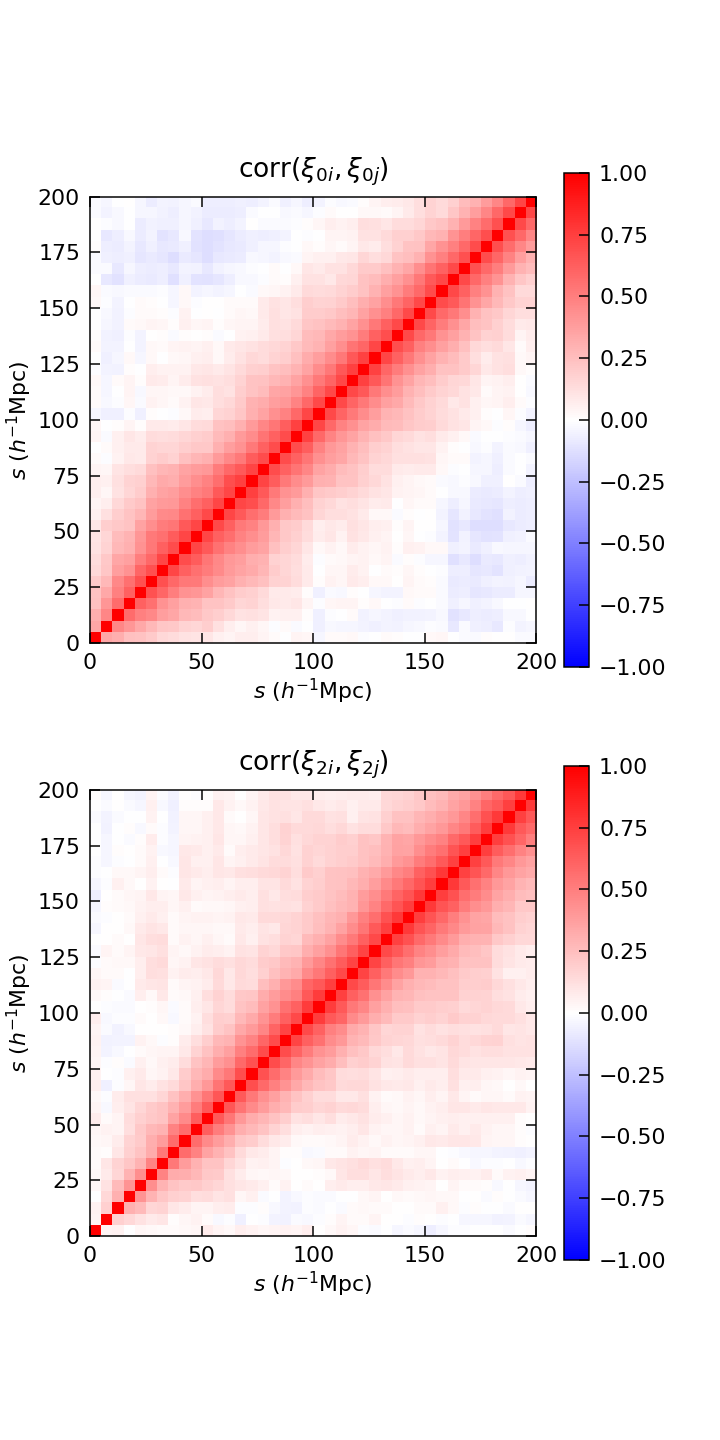}
\caption{Correlation matrix of the GLAM-QPM mocks. The upper panel shows the correlation matrix of the monopole, and the lower panel shows correlation matrix of the quadrupole. We use 40 equally spaced bins in $s$ from $0h^{-1}$Mpc to $200h^{-1}$Mpc. }
\label{corr_matrix}
\end{center}
\end{figure}

The correlation matrix $C_{ij}$ is given by
\begin{align}
	\nonumber C_{ij} &= \textnormal{corr}(x_i,x_j) \\ 
    & = \frac{\sum_{k=1}^{n_s} (x_i^k-\mu_i)(x_j^k-\mu_j)}{\sqrt[]{\sum_{k=1}^{n_s} (x_i^k-\mu_i)^2\sum_{k=1}^{n_s}(x_j^k-\mu_j)^2}}.
\end{align}The correlation matrix of the large-scale clustering measurements of the GLAM-QPM mocks is shown in Fig. \ref{corr_matrix}. These correlation matrices are all reasonable: clustering measurements between two bins with distance separated within $\sim 20 h^{-1}\rm Mpc$ are correlated.

\begin{figure}
\begin{center}
\includegraphics[width=\columnwidth]{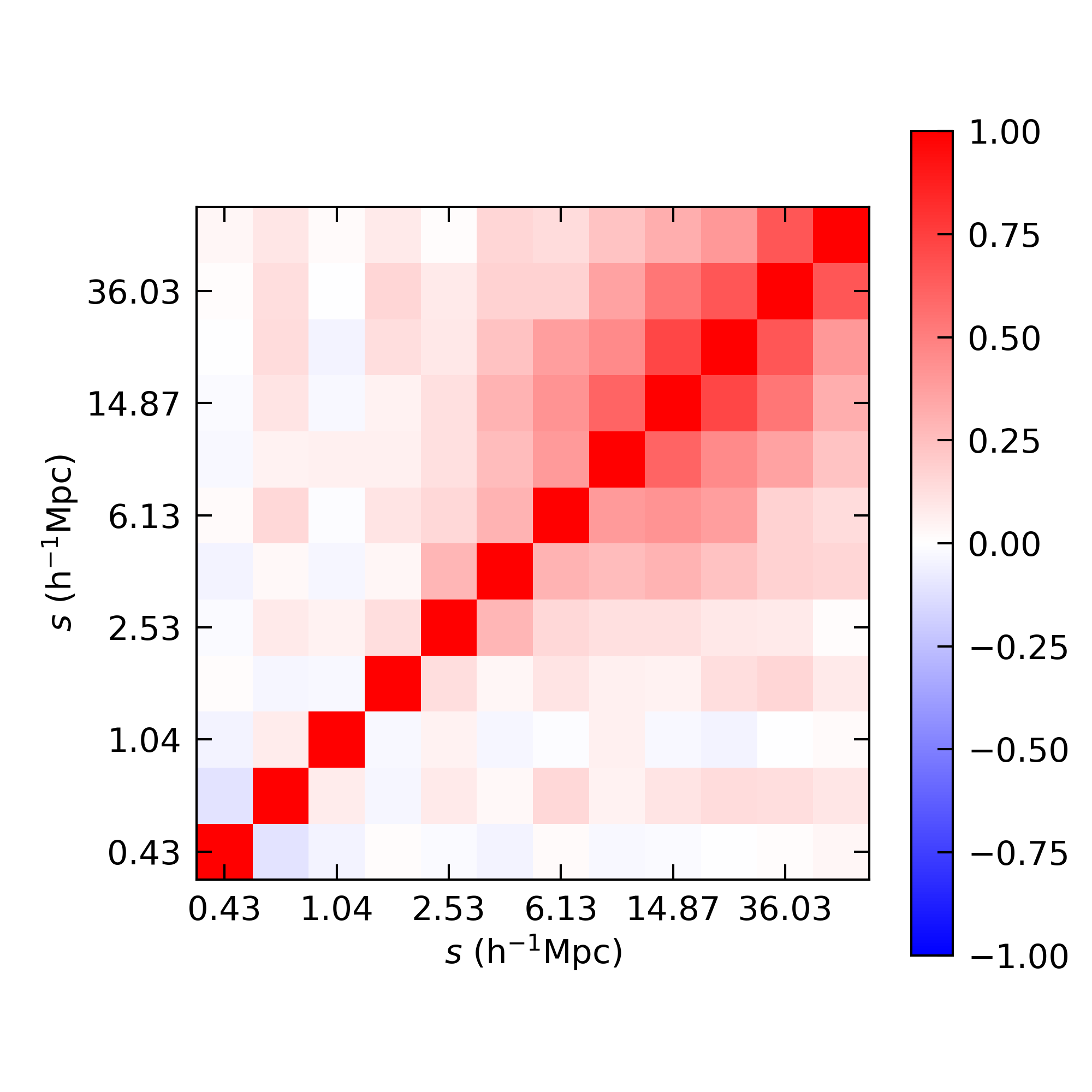}
\caption{Correlation matrix for the ELG redshift-space monopole measurements at small scales. Distance separations are binned logarithmically from $0.34h^{-1}\rm Mpc$ to $70h^{-1}\rm Mpc$.}
\label{corr_matrix_ssc}
\end{center}
\end{figure}

In Fig. \ref{corr_matrix_ssc} we show the correlation matrix for the ELG redshift space monopole measurements at small scales. The result is as expected, with a correlation between galaxy pairs on scales $s\gtrsim2.5h^{-1}\rm Mpc$, where the pairs are from two distinct halos. Monopole measurements are uncorrelated on scales $s\lesssim 2.5 h^{-1}\rm Mpc$, where galaxy pairs are within the same halo and galaxies are randomly sampled from the halo density profile.

\subsection{NGC and SGC difference}
\label{subsec: NGC_SGC_diff}
We use the GLAM-QPM mocks produced in this paper to test the consistency of the NGC and SGC measurements. For the redshift-space monopole at small scales, we measure the cross-$\chi^2=(\xi_{0_{\rm SGC}} - \xi_{0_{\rm NGC}})^TC_{\rm tot}^{-1}(\xi_{0_{\rm SGC}} - \xi_{0_{\rm NGC}})$ with 12 data points between $0.34h^{-1}\textrm{Mpc}$ and $70h^{-1}\textrm{Mpc}$, where the covariance matrix $C_{\rm tot}=C_{\rm NGC} + C_{\rm SGC}$ is estimated using the GLAM-QPM mocks. We measure $\chi^2/dof=22.3/12$ and conclude that on small scales, the SGC and NGC clustering measurements are compatible.

 In order to see whether the  difference between SGC and NGC at large scales is significant, we compute the cross-$\chi^2 $ for both the monopole and the quadrupole measurements. We choose 10 linear $s$ bins from $77.5h^{-1}\rm Mpc$ to $122.5h^{-1}\rm Mpc$, since we're most interested in the BAO scale at $s\sim 100h^{-1}\rm Mpc$. The results are $\chi^2/dof=12.53/10$ for monopoles and $\chi^2/dof=14.91/10$ for quadrupoles. The corresponding p-value is $0.25$ and $0.14$, meaning that the difference is insignificant and we cannot reject the null hypothesis that the difference between the SGC and NGC is caused by cosmic variance. We also build the $\chi^2$ distribution from our mock sample by selecting 200 NGC mocks and 200 SGC mocks (all from different GLAM simulations) and building the sample distribution from 40,000 $\chi^2$ values from each SGC-NGC pair. Our sample distribution agrees with the $\chi^2$ distribution with $dof=10$ perfectly well, indicating that our covariance matrix are valid for robust BAO and RSD analysis, as is done in \citet{arnaud2020}.

\begin{figure*}
\begin{center}
\includegraphics[width=0.95\textwidth]{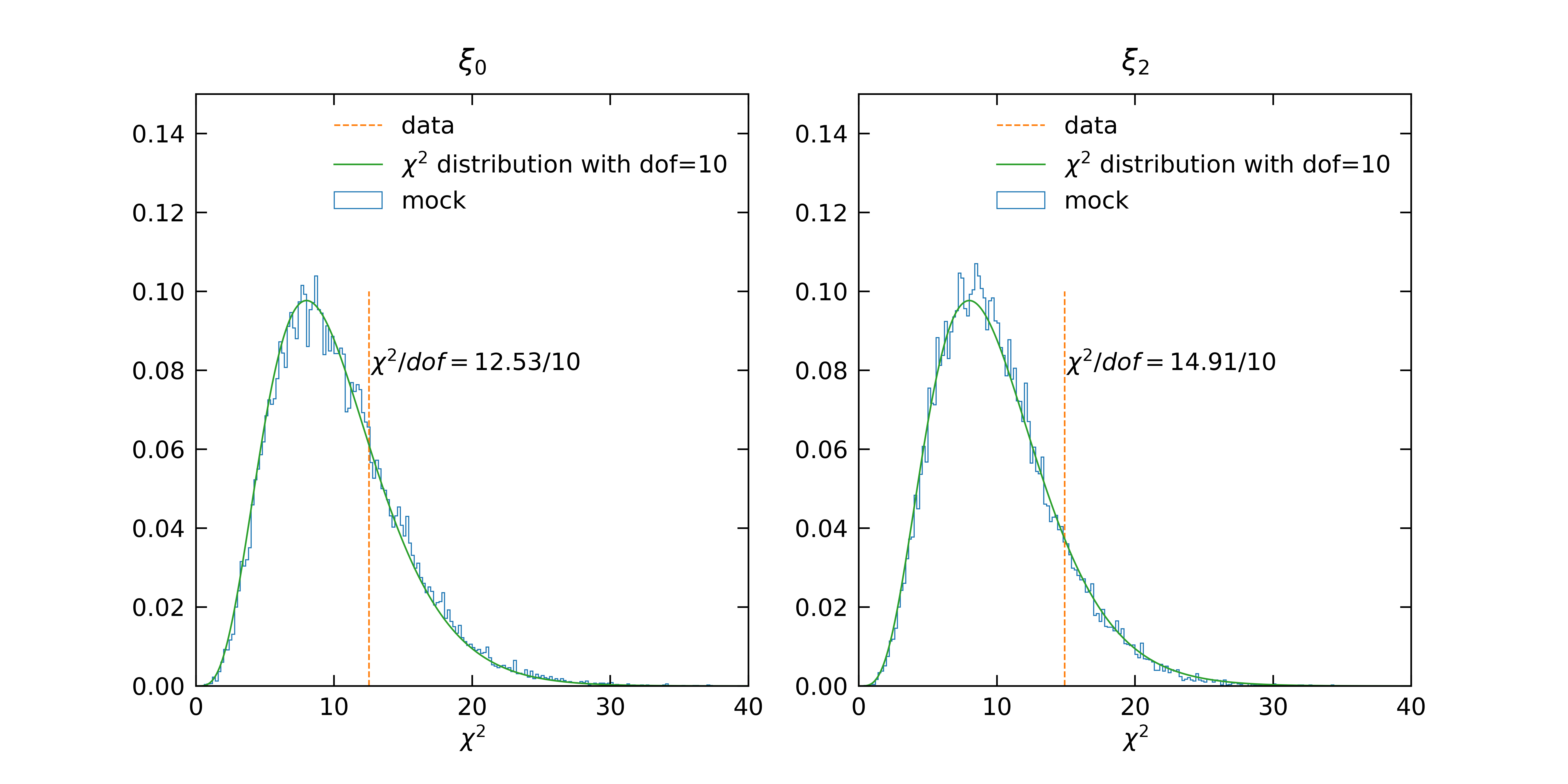}
\caption{cross-$\chi^2$ value from the SGC and NGC clustering measurements (vertical line), calculated using the covariance matrix from the GLAM-QPM mocks, given by $\chi^2= (\xi_{\rm SGC} - \xi_{\rm NGC})^TC_{\rm tot}^{-1}(\xi_{\rm SGC} - \xi_{\rm NGC})$. The left panel shows the $\chi^2$ for the monopole and the right panel is for the quadrupole. We choose 10 $r$ bins from $77.5h^{-1}\rm Mpc$ to $122.5h^{-1}\rm Mpc$ to study the clustering signal around BAO peak $(r\sim 100h^{-1}\rm Mpc)$. We also show the $\chi^2$ distribution from 40,000 pairs of NGC and SGC mocks (blue histogram), in order to see where the data falls in the distribution. A $\chi^2$ distribution with $dof = 10$ is shown by the green curve. }
\label{chi2}
\end{center}
\end{figure*}

\section{Summary}
We present 2000 GLAM-QPM mock catalogs for the eBOSS DR16 ELG sample. We use GLAM simulations to quickly and accurately produce the dark matter density field and the QPM method to assign dark matter halos to particles in the simulation. The haloes are then populated with ELGs using a HOD methodology. We have calibrated the HOD parameters for the eBOSS ELG sample to model the large-scale bias of ELGs. The majority of central galaxies falls in halos with mass between $10^{11}h^{-1}\rm M_\odot$ and $10^{13}h^{-1}\rm M_\odot$, and the satellite fraction of our HOD model is $17.4\%$. The eBOSS ELG survey geometry and radial selection functions are applied to our mocks. This set of mock catalogs is used in the eBOSS ELG RSD analysis in \citet{arnaud2020}.

We've shown that the GLAM-QPM mock catalogs agree with ELG data at large scales, in general, within $2\sigma$ for monopole. For quadrupole the ELG SGC data shows higher clustering signal than GLAM-QPM mocks. We examined the cross-$\chi^2$ value for SGC and NGC around BAO scales, and find $\chi^2/dof=10.09/10$ for the monopole and $\chi^2/dof=14.86/10$ for the quadrupole. We cannot conclude that the difference between SGC and NGC are due to reasons other than cosmic variance. 

The GLAM-QPM mock galaxy catalogs will be published along with eBOSS DR16 galaxy catalog release, and will also be placed at the Skies and Universes site \citep{sky_universe} \footnote{www.skiesandunierses.org}.

\section*{acknowledgments}

SL is grateful to the support from CCPP, New York University. J.L.T. and M.R.B. are supported by NSF Award 1615997. F.P. and A.K. acknowledges support from the Spanish MICINU grant GC2018-101931-B-100. G.R. acknowledges support from the National Research Foundation of Korea (NRF) through Grants No. 2017R1E1A1A01077508 and No. 2020R1A2C1005655 funded by the Korean Ministry of Education, Science and Technology (MoEST), and from the faculty research fund of Sejong University. The GLAM simulations used in this paper were done on MareNostrum-4  at the Barcelona Supercomputer Center in Spain. 

Funding for the Sloan Digital Sky Survey IV has been provided by
the Alfred P. Sloan Foundation, the U.S. Department of Energy Office of
Science, and the Participating Institutions. SDSS-IV acknowledges
support and resources from the Center for High-Performance Computing at
the University of Utah. The SDSS web site is www.sdss.org.

SDSS-IV is managed by the Astrophysical Research Consortium for the 
Participating Institutions of the SDSS Collaboration including the 
Brazilian Participation Group, the Carnegie Institution for Science, 
Carnegie Mellon University, the Chilean Participation Group, the French Participation Group, Harvard-Smithsonian Center for Astrophysics, 
Instituto de Astrof\'isica de Canarias, The Johns Hopkins University, 
Kavli Institute for the Physics and Mathematics of the Universe (IPMU) / 
University of Tokyo, Lawrence Berkeley National Laboratory, 
Leibniz Institut f\"ur Astrophysik Potsdam (AIP),  
Max-Planck-Institut f\"ur Astronomie (MPIA Heidelberg), 
Max-Planck-Institut f\"ur Astrophysik (MPA Garching), 
Max-Planck-Institut f\"ur Extraterrestrische Physik (MPE), 
National Astronomical Observatory of China, New Mexico State University, 
New York University, University of Notre Dame, 
Observat\'ario Nacional / MCTI, The Ohio State University, 
Pennsylvania State University, Shanghai Astronomical Observatory, 
United Kingdom Participation Group,
Universidad Nacional Aut\'onoma de M\'exico, University of Arizona, 
University of Colorado Boulder, University of Oxford, University of Portsmouth, 
University of Utah, University of Virginia, University of Washington, University of Wisconsin, 
Vanderbilt University, and Yale University.

\bibliographystyle{mnras}
\bibliography{main}


\bsp	
\label{lastpage}
\end{document}